\documentclass[fleqn,usenatbib,useAMS]{mnras}

\usepackage{ae,aecompl}
\usepackage{graphicx}
\usepackage{amsmath,amsfonts,amssymb}
\usepackage[T1]{fontenc}

\newcommand{\nuc}[2]{${}^{#2} \rm #1$}
\newcommand{\Ms}{M_{\odot}}
\newcommand{\K}{ {\,\rm K} }

\title[Impact of isomers on a kilonova]
{The impact of isomers on a kilonova associated with neutron star mergers}

\author[S. Fujimoto and M. Hashimoto]{
Shin-ichiro Fujimoto$^{1}$\thanks{E-mail: fuji@kumamoto-nct.ac.jp},
Masa-aki Hashimoto$^{2}$
\\
$^{1}$National Institute of Technology, Kumamoto College, Kumamoto 861-1102, Japan
\\
$^{2}$Department of Physics, Kyushu University, Fukuoka 819-0395, Japan \\
}

\date{Accepted XXX. Received YYY; in original form ZZZ}

\pubyear{2019}

\begin{document}
\label{firstpage}
\pagerange{\pageref{firstpage}--\pageref{lastpage}}
\maketitle

\begin{abstract}
We investigate the significance of isomers on a kilonova associated with neutron star mergers (NSMs) for the first time.
We calculate the evolution of abundances and energy generation rates ($\dot{\epsilon}_{\rm nuc}$) of ejecta from NSMs,
taking into account $\beta^-$ decay through isomers.
We find that for ejecta with electron fraction ($Y_e$) of $0.2-0.3$, 
$\dot{\epsilon}_{\rm nuc}$ is appreciably changed from those without isomers, due to the large change in timing of $\beta^-$ decay through the isomeric states 
of \nuc{Sn}{123,125,127} and \nuc{Sb}{128}.
In particular, the effects of the isomers on $\dot{\epsilon}_{\rm nuc}$ are prominent for ejecta of $Y_e \sim 0.25$,
which could emit a fraction of an early, blue component of a kilonova observed in GW170817.
When the excitation from a ground state to its isomeric state is important, 
the isomers of \nuc{Sb}{129} and \nuc{Te}{129, 131} also cause additional and appreciable change in $\dot{\epsilon}_{\rm nuc}$.
Furthermore, we demonstrate that larger amounts of lanthanide-free ejecta result in a better fit of
an observed light curve of the kilonova in GW170817 if the isomers are taken into account.
\end{abstract}




\section{Introduction}\label{sec:intro} 

A gravitational wave source, GW170817, has been identified to be a neutron star merger (NSM)~\citep{2017PhRvL.119p1101A}
and the association was observationally confirmed with a kilonova, AT2017gfo, 
whose light curve has been modeled with two components; early blue and later red ones~\citep{2017ApJ...848L..17C, 2017PASJ...69..102T, 2017ApJ...848L..27T, 2017Natur.551...80K, 2017ApJ...851L..21V, 2017Natur.551...75S}.
In the scenario, the kilonova is possibly powered by the radioactive decay of neutron($n$)-rich matter freshly synthesized in ejecta from NSM 
through $r$-process nucleosynthesis~\citep{1998ApJ...507L..59L, 2014ApJ...789L..39W}.
The early blue component is thought to be emitted from post-merger ejecta, which is mildly $n$-rich (electron fraction, $Y_e$, $\ge 0.25-0.3$), 
and thus has a small amount of lanthanide, whose opacity is much larger than Fe.
Fitting of the model to the light curve requires the post-merger ejecta 
to be massive ($> 0.01 \Ms$) and fast ($> 0.1c$)~\citep{2017ApJ...848L..17C, 2017ApJ...848L..27T, 2017ApJ...851L..21V}, where $c$ is the speed of light.
On the other hand, the later red component corresponds to the emission from dynamical ejecta, 
which is ejected via tidal interactions between two neutron stars, and whose $Y_e$ is less than $0.2-0.25$
resulting in a large amount of lanthanide and very high opacity in the ejecta.
The dynamical ejecta are required to be heavier ($\sim 0.05 \Ms$) and faster (around $0.2c$) 
than the post-merger ejecta~\citep{2017ApJ...848L..17C, 2017ApJ...848L..27T, 2017ApJ...851L..21V}.

There exist nuclei whose isomer, or a quasi-stable state of the nuclei, 
has a branch of the $\beta^-$ decay to their daughter nuclei and the $\beta^-$ decay half-life largely different from that of their ground states (GSs).
If GS of such a nuclide excites to its isomeric state (IS) or IS is preferentially populated over GS~\footnote{
This is the case for the isomer of \nuc{Sn}{123,125,127} and \nuc{Sb}{128}, as shown in Appendix A}
during the ejection from NSMs, 
a light curve of the kilonova is likely to be modified due to the large change in the half lives of the nuclei through ISs.
In particular, the influences of isomers are possibly significant for unstable nuclei whose mass number is around 120-140 
and which decay to second-peak nuclei,
since the nuclei are primary heating sources of the kilonova~\citep{2010MNRAS.406.2650M}.

In the present {\it Letter}, we investigate the impact of isomers on a kilonova from NSMs.
Taking account of isomers, 
we calculate evolution of abundances and energy generation rates, $\dot{\epsilon}_{\rm nuc}$, of ejecta from NSMs.

This {\it Letter} is organized as follows. In section 2, we select nuclei whose isomer is taken into account in the light of observational importance of a kilonova
and we briefly present properties of the selected nuclei.
We show our results of the evolution of abundances and $\dot{\epsilon}_{\rm nuc}$ of ejecta from NSMs in section 3, and fit an observed light curve with our models 
varying masses and velocities of the ejecta in section 4.
We discuss our results in section 5 and summarize our results in section 6.

\section{Isomers}\label{sec:isomers} 

In order to evaluate effects of isomers on a kilonova light curve, 
we select nuclei whose isomer is taken into account in the light of observational importance, 
so that they satisfy all the following four criteria;
(1) nuclei with non-zero, $\beta^-$ decay branching-ratio (BR) of IS:
(2) nuclei with $\beta^-$ decay half-lives of GS or IS from $\sim$ 0.5 hours to 200 days
and the half life of IS differs from that of GS by a factor of $\ge 10$: 
(3) nuclei with an isomer whose excitation energy from the GS to the IS is less than 2 MeV: 
and (4) nuclei whose mass number, $A$, is $120-140$.
Here we limit an excitation energy of isomers below 2 MeV, because temperatures of ejecta decrease from an initial temperature, or $9 \times 10^9 \K$ (Sec. 3).
We find nine nuclei, \nuc{Sn}{121,123,125,127}, \nuc{Sb}{128,129}, \nuc{Te}{129,131}, and \nuc{I}{134}, which satisfy the four criteria.
For \nuc{Sn}{123, 125, 127}, \nuc{Sb}{129}, and \nuc{I}{134},
a $\beta^-$ decay half-live of IS is much shorter than that of GS,
while the half life is much longer for \nuc{Sn}{121}, \nuc{Te}{129, 131}.
It should be noted that the ISs of \nuc{Sn}{123,125,127}, and \nuc{Sb}{128} dominate over their GSs,
while the isomers of the others can be ignored, even when IS is not excited from GS (Appendix \ref{sec:appendix}).

\section{Evolution of abundances and energy generation rates of ejecta}\label{sec:abundances} 

We adopt a simplified dynamical model~\citep{2017CQGra..34j4001R} of the ejecta, which is assumed to be uniformly expanded with a constant velocity, 
to follow the evolution of density of the ejecta.
Temperatures of the ejecta are evaluated with an equation of state~\citep{1996ApJS..106..171B}, 
taking into account the energy generation through nuclear reactions~\citep{2017CQGra..34j4001R}.
The model has four parameters; the mass of the ejecta ($M_{\rm ej}$), the expansion velocity ($v_{\rm ex}$), the electron fraction ($Y_e$), and entropy per baryon ($s$),
where $Y_e$ and $s$ are evaluated when the temperature is equal to $9 \times 10^9 \rm K$.
We have performed calculations for parameters of $Y_e=0.1, 0.15, 0.2, 0.25$, and 0.3, and $v_{\rm ex}/c = 0.05, 0.1$, and 0.2 
with fixed parameters $M_{\rm ej} = 0.03M_\odot$, which is comparable to masses adopted in kilonova models for GW170817~\citep{2017PASJ...69..102T, 2017Natur.551...80K},
and $s=10 k_{\rm B}$, where $k_{\rm B}$ is the Boltzmann constant.
We investigate the following three cases;
(a) the nine nuclei (\nuc{Sn}{121,123,125,127}, \nuc{Sb}{128,129}, \nuc{Te}{129,131}, and \nuc{I}{134}) are assumed to be always stayed in their GSs:
(b) the nine nuclei in their ISs with 100\% $\beta^-$ BRs: and 
(c) the four nuclei, \nuc{Sn}{123,125,127}, and \nuc{Sb}{128}, whose IS largely dominates over its GS even when IS is not excited from GS (Appendix \ref{sec:appendix}), 
are assumed to be always stayed in their ISs with 100\% $\beta^-$ BRs.
Here the $\beta^-$ BR is less than 100\% except for \nuc{Sn}{123, 125, 127}.
It should be emphasized that for case (b) effects of the isomers are the maximum on an energy generation rate 
and that case (c) corresponds to a rough approximation to a situation without the excitation from GS to IS.

We follow abundance and temperature evolution of the ejecta with a nuclear reaction network, which includes 4070 nuclei from $n$, $p$ to Fm (atomic number, $Z$, = 100)
and is employed in our previous works for $r$-process nucleosynthesis~\citep{2007ApJ...656..382F, 2008ApJ...680.1350F}.
We note that in the network we adopt a large number of reaction rates theoretically evaluated with FRDM~\citep{1995ADNDT..59..185M}.
We assume that reaction rates of IS other than $\beta^-$ decays are same as those of GS for nuclei in which isomer is taken into account.
We change half lives via the $\beta^-$ decay for the nine nuclei according to cases (a), (b), and (c).
The temperatures are found to be slightly higher for slower ejecta but weakly depend on $Y_e$ and the effects of the isomers.

\begin{figure}
 \begin{minipage}{1.0\hsize}
  \includegraphics[scale=0.65]{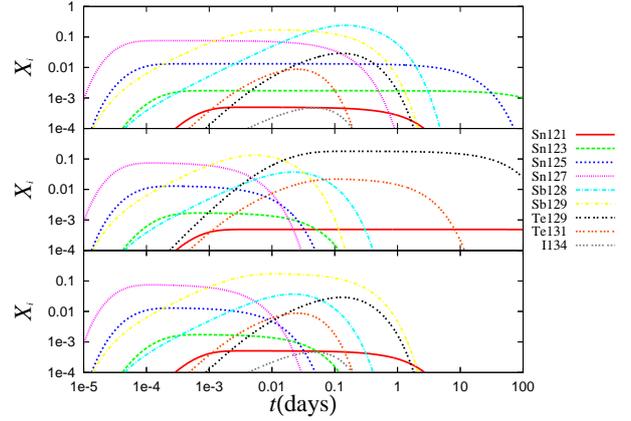}
  \vspace*{-10pt}
  \caption{Evolution of mass fractions, $X_i$, of the nine nuclei for cases (a) (top panel), (b) (middle panel), and (c) (bottom panel) for $Y_e = 0.25$ and $v_{\rm ex}=0.05c$.
  The evolution of the mass fractions for cases (b) and (c) is largely changed from that for case (a) due to 
  $\beta^-$ decay through isomers, whose half live is largely different from that of GSs.}
  \label{fig:abundance}
 \end{minipage}
\end{figure}

Figure \ref{fig:abundance} shows abundance evolution of the nine nuclei (\nuc{Sn}{121,123,125,127}, \nuc{Sb}{128,129}, \nuc{Te}{129,131}, and \nuc{I}{134})
for $Y_e = 0.25$ and $v_{\rm ex} = 0.05c$ and for cases (a), (b), and (c).
Evolution of the mass fractions for cases (b) and (c) is largely changed from that for case (a) because of the change of $\beta^-$ decay half-lives via the ISs.
We note that ejecta with $Y_e \ge 0.25$ are lanthanide-free and nuclei with $A = 120-140$ are abundantly produced in ejecta with $Y_e = 0.2-0.3$,
and that the nine nuclei are abundantly produced through a sequence of $\beta^-$ decay of mildly $n$-rich nuclei synthesized via $r$-process.

\begin{figure}
 \begin{minipage}{1.0\hsize}
  \includegraphics[scale=0.65]{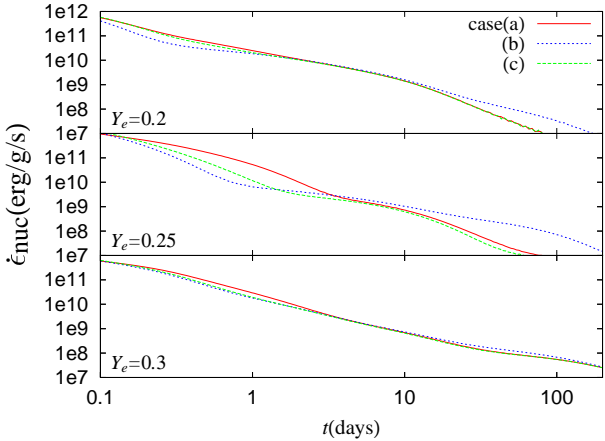}
  \vspace*{-5pt}
  \caption{
  Evolution of $\dot{\epsilon}_{\rm nuc}$ for $v_{\rm ex}=0.05c$ and for $Y_e=$ 0.2 (top panel), 0.25 (middle panel), and 0.3 (bottom panel).
  Due to the large differences in $\beta^-$ decay half-lives between GS and IS, 
  $\dot{\epsilon}_{\rm nuc}$ for case (b) (dotted lines) decrease at the early phase ($<$ a few days) 
  but increase after a few days compared with case (a) (solid lines).
  Differences between cases (a) and (c) (dashed lines) are smaller than those between cases (a) and (b).
  }
  \label{fig:enuc-Ye}
 \end{minipage}
\end{figure}

Figure \ref{fig:enuc-Ye} shows evolution of $\dot{\epsilon}_{\rm nuc}$ for $v_{\rm ex}=0.05c$ and cases (a), (b), and (c).
We find that there are appreciable differences in $\dot{\epsilon}_{\rm nuc}$ between cases (a) and (b) for $Y_e=0.2-0.3$.
In particular, the differences in $\dot{\epsilon}_{\rm nuc}$ are prominent for the $Y_e = 0.25$ ejecta.
Note that the ejecta with $Y_e \sim 0.25$ have physical properties similar to those of material that emits an early, blue component of a kilonova
observed in GW170817~\citep{2017PASJ...69..102T}.
We also find that the differences between cases (a) and (c) are appreciable but smaller than those between cases (a) and (b).

At the early phase, $t = (0.1-3)$ days, $\dot{\epsilon}_{\rm nuc}$ in case (b) are smaller that those in case (a)
due to the earlier $\beta^-$ decays through the isomers of \nuc{Sn}{127} and \nuc{Sb}{129} and later $\beta^-$ decays via the isomer of \nuc{Te}{129},
while the earlier $\beta^-$ decay of the isomer of \nuc{Sb}{128} compensates a fraction of the decrease (Fig. \ref{fig:abundance}).
The differences in $\dot{\epsilon}_{\rm nuc}$ between cases (a) and (c) are smaller than those between cases (a) and (b),
since the isomers of \nuc{Sb}{129} and \nuc{Te}{129} are not included in case (c).

During the later phase, $t = (3-10)$ days, 
due to the earlier $\beta^-$ decay through the isomer of \nuc{Sb}{128}, $\dot{\epsilon}_{\rm nuc}$ decreases in case (c) compared with that in case (a),
while $\dot{\epsilon}_{\rm nuc}$ in case (b) is comparable to that in case (a).
For case (b), the decrease in $\dot{\epsilon}_{\rm nuc}$ due to the isomer of \nuc{Sb}{128} 
is covered with the increase in $\dot{\epsilon}_{\rm nuc}$ via the later $\beta^-$ decay through the isomer of \nuc{Te}{131}, which is not taken into account in case (c).

For $t = (10-100)$ days, $\dot{\epsilon}_{\rm nuc}$ increases in case (b) compared with that in case (a)
due to the later $\beta^-$ decay through the isomer of \nuc{Te}{129}, 
although a fraction of the increase due to \nuc{Te}{129} is compensated with the earlier $\beta^-$ decay of the isomer of \nuc{Sn}{125},
which causes the decrease in $\dot{\epsilon}_{\rm nuc}$ in case (c) compared with that in case (a).

\begin{figure}
 \begin{minipage}{1.0\hsize}
  \includegraphics[scale=0.65]{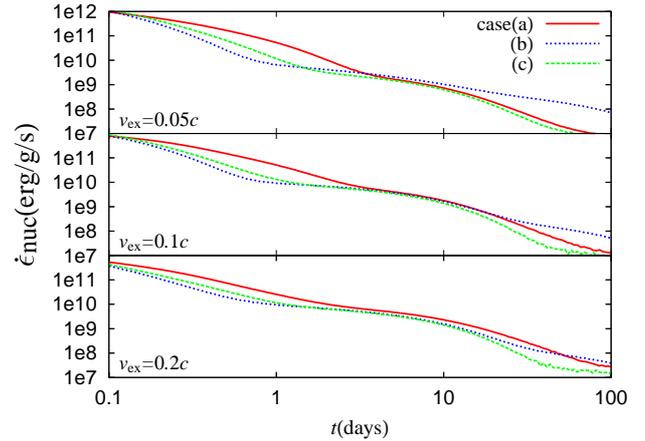}
  \vspace*{-5pt}
  \caption{
  Evolution of $\dot{\epsilon}_{\rm nuc}$ for $Y_e=0.25$ and for $v_{\rm ex}/c=$ 0.05 (top panel), 0.1 (middle panel), and 0.2 (bottom panel).
  Differences in $\dot{\epsilon}_{\rm nuc}$ between cases (a) (solid lines) and (b) (dotted lines) and between cases (a) and (c) (dashed lines) 
  are larger for slower ejecta due to higher temperatures.
  }
  \label{fig:enuc-Vex}
 \end{minipage}
\end{figure}

We find that $\dot{\epsilon}_{\rm nuc}$ depends not only on $Y_e$ but also on $v_{\rm ex}$.
Figure \ref{fig:enuc-Vex} shows evolution of $\dot{\epsilon}_{\rm nuc}$ for $Y_e=$ 0.25 with $v_{\rm ex}/c=0.05$, 0.1, and 0.2, and for cases (a), (b), and (c).
The distinctions in $\dot{\epsilon}_{\rm nuc}$ between cases (a) and (b) and between cases (a) and (c) are larger for slower ejecta, due to higher temperatures.

\vspace*{-12pt}

\section{Fitting to a light curve of the kilonova observed in GW170817}\label{sec:obs}

\begin{figure}
 \begin{minipage}{1.0\hsize}
  \includegraphics[scale=0.8]{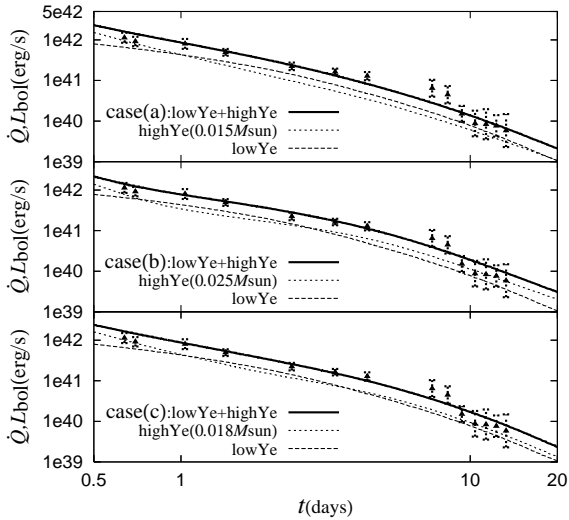}
  \vspace*{-20pt}
  \caption{
  Evolution of $\dot{Q}$ for the two-component ejecta (solid lines), that for the low-$Y_e$ component (dashed lines), and that for the high-$Y_e$ component (dotted lines), 
  and for case (a) with $M_{\rm ej, high} = 0.015 \Ms $ (top panel), 
  case (b) with $M_{\rm ej, high} = 0.025 \Ms $ (middle panel),
  and case (c) with $M_{\rm ej, high} = 0.018 \Ms $  (bottom panel).
  Larger amounts of the high-$Y_e$, or lanthanide-free ejecta are required to fit $L_{\rm bol}$ observed in GW170817
  (filled triangles with an error bar), 
  if we take account of the isomers (cases (b) and (c)).
  }
  \label{fig:luminosity}
 \end{minipage}
\end{figure}

We examine impact of the isomers on a light curve of a kilonova in NSMs.
We evaluate generation rates of thermal energy as, $\dot{Q} = f_{\rm th} \, \dot{\epsilon}_{\rm nuc} \, M_{\rm ej}$, 
where $f_{\rm th}$ is a fraction of thermal energy to the liberated energy via nuclear reactions~\citep{2016ApJ...829..110B, 2017CQGra..34j4001R}.
We construct a simplified two-component emission model, in which low and high-$Y_e$ ejecta are taken into account, to evaluate $\dot{Q}$.
We assume that the low-$Y_e$ ejecta contains contributions with $Y_e = $ 0.1, 0.15, and 0.2 and high-$Y_e$ ejecta with $Y_e = $ 0.25, 0.3, 0.35, and 0.4 
with a uniform mass distribution.
The mass and velocity of the low-$Y_e$ ejecta and the velocity of the high-$Y_e$ ejecta are fixed to be $0.03\Ms$, $0.1c$, and $0.2c$,
while the mass of the high-$Y_e$ ejecta, $M_{\rm ej, high}$, is adequately set to fit bolometric luminosities, $L_{\rm bol}$, observed in GW170817~\citep{2017Natur.551...75S}.
Note that the high-$Y_e$ component is lanthanide-free while lanthanide is appreciable in the low-$Y_e$ component.
The high and low-$Y_e$ ejecta roughly mimic the early blue and later red components of the kilonova observed in GW170817, respectively.
Figure \ref{fig:luminosity} shows evolution of $\dot{Q}$ in units of $\rm erg\,s^{-1}$ 
for cases (a), (b) and (c) (top, middle, and bottom panels).
Heavier high-$Y_e$ components are found to be required to fit $L_{\rm bol}$ for cases (b) and (c) compared with case (a).
We note that at $t = (2-8)$ days, $\dot{Q}$ is smaller and the fitting becomes worse if we adopt the velocity of the high-$Y_e$ ejecta of $0.05c$. 

We note that the adopted mass of the low-$Y_e$ ejecta ($0.03\Ms$) and velocities of the high-$Y_e$ ejecta ($0.1c$) are higher and faster than 
those estimated with recent multi-dimensional (D) simulations of NSMs~(e.g. \citet{2015MNRAS.448..541J, 2016MNRAS.460.3255R, 2016PhRvD..93l4046S, 2018ApJ...860...64F})
and that at $t > 3$ days, the contribution to $\dot{Q}$ of the high-$Y_e$ ejecta dominate that of the low-$Y_e$ ejecta, 
in spite of the correspondence of the high- and low-$Y_e$ ejecta to the {\itshape early} blue and {\itshape later} red components, respectively.
Mass fractions of lanthanides, $X_{\rm lan}$, are estimated to be 0.019, 0.016, and 0.018 for cases (a), (b), and (c), which are slightly higher than 
an upper bound ($\sim 0.01$)~\citep{2017Natur.551...80K},
although $X_{\rm lan}$ becomes below the upper bound if we adopt a smaller mass of the low-$Y_e$ ejecta ($< 0.01\Ms$), consistent with the multi-D simulations of NSMs.
We should stress that our fitting of $\dot{Q}$ to $L_{\rm bol}$ is based on the simplified dynamical and two-component emission models.
Multi-D models with a lower mass of the low-$Y_e$ ejecta~\citep{2018ApJ...865L..21K} and with the isomers
might lead to the consistency among the observation of the kilonova and the results of the multi-D simulations.

\section{Discussion}\label{sec:discussion} 

\subsection{Isomers of nuclei with $A < 120$ or $A > 140$}

We have selected nuclei whose isomer is taken into account in terms of the four criteria (Sec. 2).
We investigate effects of IS of nuclei with $A < 120$ or $A > 140$.
We adopt additional five nuclei, \nuc{Rh}{106}, \nuc{Cd}{115}, \nuc{Lu}{177}, \nuc{Ir}{194}, and \nuc{Au}{200}, 
in terms of the criteria (1), (2), and (4). 
We have calculated the evolution of abundances and $\dot{\epsilon}_{\rm nuc}$
for $Y_e=0.1-0.3$ and $v_{\rm ex} = 0.05c$,
assuming that 
the 14 nuclei (\nuc{Sn}{121,123,125,127},\nuc{Sb}{128,129},\nuc{Te}{129,131}, \nuc{I}{134}, \nuc{Rh}{108}, \nuc{Cd}{115}, \nuc{Lu}{177}, \nuc{Ir}{194}, and \nuc{Au}{200})
are always stayed in their ISs.
We find that change in $\dot{\epsilon}_{\rm nuc}$ is negligible due to IS of nuclei with $A < 120$ or $A > 140$ except for ejecta with $Y_e = 0.1$,
although the change is small even for $Y_e = 0.1$.

\vspace*{-12pt}

\subsection{Excitation from a ground state to the isomeric state}

We briefly discuss the excitation from GS to IS of the nine nuclei.
Although we neglect the excitation from GS to IS, instead we simply consider the three limiting cases (a) to (c) in the present study,
the excitation becomes important when the temperature of ejecta is comparable to the excitation energy of the isomers, $E_{\rm ex}$.
Since the temperature is less than 0.01MeV when the nine nuclei are abundantly produced through a sequence of $\beta^-$ decay (>0.001 days (Fig. 1)),
the excitation from GS to IS of nuclei with larger $E_{\rm ex}$, or 
\nuc{Sb}{129}, \nuc{Te}{129}, \nuc{Te}{131} and \nuc{ I}{134}, whose $E_{\rm ex}$ are 1.85, 0.11, 0.18, and 0.32 MeV, 
is negligible unless an extra heating operates in the ejecta in addition to the nuclear heating.
We note that the temperature of the ejecta might increase via internal shock and/or shock induced through interactions between the ejecta and circum-stellar matter.
On the other hand, for nuclei with smaller $E_{\rm ex}$, or \nuc{Sn}{121,123,125,127} and \nuc{Sb}{128},
whose $E_{\rm ex}$ are 0.006, 0.025, 0.028, 0.0047 and 0.02 MeV, the excitation could be important
and result in $\beta^-$ decay of \nuc{Sn}{121} at a very late phase through the isomer whose half life is 44 years.
More quantitative evaluation of effects of the excitation on $\dot{\epsilon}_{\rm nuc}$ and the light curve requires more detailed network calculation,
in which excited states are taken into account in addition to GSs and ISs of the nine nuclei, as in case for \nuc{Al}{26}~\citep{2011ApJS..193...16I} but much more complicated.
The calculation is beyond the scope of the present study. We postpone the calculation to a future paper.

\vspace*{-12pt}

\section{Concluding remarks}\label{sec:conclution} 

In this {\it Letter}, 
we have investigated the impact of isomers on the light curve of a kilonova associated with NSMs for the first time.
Taking into account isomers of the nine nuclei, 
\nuc{Sn}{121,125,123,127}, \nuc{Sb}{128,129}, \nuc{Te}{129,131}, and \nuc{I}{134}, 
which are selected in the light of observational importance in a kilonova associated with NSMs,
we calculated the evolution of abundances and energy generation rates, $\dot{\epsilon}_{\rm nuc}$, of the ejecta, 
whose dynamics is evaluated with a simplified model with four parameters; $M_{\rm ej}$, $v_{\rm ex}$, $Y_e$, and $s$.
We have performed the calculations for three limiting cases;
(a) all nuclei always stay in GS, 
(b) the above nine nuclei always stay in IS, 
and (c) the four nuclei, \nuc{Sn}{123,125,127} and \nuc{Sb}{128}, in IS.
We study ejecta for parameters of $Y_e=0.1, 0.15, 0.2, 0.25$, and 0.3, and $v_{\rm ex}/c = 0.05, 0.1$, and 0.2 
with fixed parameters $M_{\rm ej} = 0.03M_\odot$ and $s=10 k_{\rm B}$.
Here we note that \nuc{Sn}{125,123,127} and \nuc{Sb}{128} are shown to stay in IS dominantly, even when GS is not excited to IS,
from careful examination of the decay schemes of isobars of the nuclei (Appendix A).
We emphasized that case (c) corresponds to a rough approximation to a situation without the excitation from GS to IS.

We find that for $Y_e= 0.2, 0.25$, and 0.3, 
$\dot{\epsilon}_{\rm nuc}$ is appreciably changed from that without isomers, due to the change in timing of $\beta^-$ decay through ISs of \nuc{Sn}{123,125,127} and \nuc{Sb}{128}.
If the excitation from GS to IS is important, the isomers of \nuc{Sb}{129} and \nuc{Te}{129, 131} also cause additional changes in $\dot{\epsilon}_{\rm nuc}$.
In particular, the impact of the isomers on $\dot{\epsilon}_{\rm nuc}$ is significant for $Y_e= 0.25$ ejecta,
which could correspond to a fraction of the early, blue component of a kilonova in GW170817,
because the ejecta have abundant nuclei around the second-peak with $A=120-140$ and negligible lanthanide.
We also find that the effects in $\dot{\epsilon}_{\rm nuc}$ are minor for IS of nuclei with $A < 120$ or $A > 140$.
Finally, employed with the two-component emission model, 
we demonstrate that larger amounts of lanthanide-free ejecta are required to fit an observed light curve of the kilonova in GW170817 if the isomers are taken into account.

\vspace*{-12pt}

\section*{Acknowledgements}
We thank Dr. H. Miyatake and Dr. T. Hayakawa for helpful information on isomers
and the anonymous referee for detailed comments that helped us to improve our manuscript.




\vspace*{-12pt}

\input{ms.bbl}

\vspace*{-12pt}

\appendix
\section{Effective branching ratios of $\beta^-$ decaying isomers}
\label{sec:appendix} 

For the nine isomers, \nuc{Sn}{121,123,125,127}, \nuc{Sb}{128,129}, \nuc{Te}{129,131}, and \nuc{I}{134}, 
we evaluate an effective $\beta^-$ BR, which is defined as a fraction of isomers decaying to its daughter nuclei via $\beta^-$ decay
when a GS of the nuclei is not excited to its IS,
using nuclear data taken from NUDAT2~\footnote{https://www.nndc.bnl.gov/nudat2/}.

\subsection{$A = 121$ isobars}

There are three decay sequences to \nuc{Sn}{121}${}^m$ from \nuc{Ag}{121}, which has no isomer with a $\beta^-$ branch; 
(a) \nuc{Ag}{121} $\rightarrow$ \nuc{Cd}{121}${}^g$ $\rightarrow$ \nuc{In}{121}${}^m$$\rightarrow$\nuc{Sn}{121}${}^m$,
(b) \nuc{Ag}{121}$\rightarrow$\nuc{Cd}{121}${}^g$$\rightarrow$\nuc{In}{121}${}^g$$\rightarrow$\nuc{Sn}{121}${}^m$, and 
(c) \nuc{Ag}{121}$\rightarrow$\nuc{Cd}{121}${}^m$$\rightarrow$\nuc{In}{121}${}^g$$\rightarrow$\nuc{Sn}{121}${}^m$, 
whose BRs are 40.25\%, 3.85\%, and 0.97\%.
Since $\beta^-$ BR of \nuc{Sn}{121}${}^m$ to \nuc{Sb}{121} is 22.4\%,
the effective BR to \nuc{Sn}{121}${}^m$ results in 9.88\%.

\subsection{$A = 123$ isobars}

There are four decay sequences to \nuc{Sn}{123}${}^m$ from \nuc{Ag}{123}, which has no isomer with a $\beta^-$ branch; 
(a) \nuc{Ag}{123} $\rightarrow$ \nuc{Cd}{123}${}^g$ $\rightarrow$ \nuc{In}{123}${}^g$ $\rightarrow$ \nuc{Sn}{123}${}^m$, 
(b) \nuc{Ag}{123}$\rightarrow$\nuc{Cd}{123}${}^g$$\rightarrow$\nuc{In}{123}${}^m$$\rightarrow$\nuc{Sn}{123}${}^m$, 
(c) \nuc{Ag}{123}$\rightarrow$\nuc{Cd}{123}${}^m$$\rightarrow$\nuc{In}{123}${}^g$$\rightarrow$\nuc{Sn}{123}${}^m$, and
(d) \nuc{Ag}{123} $\rightarrow$ \nuc{Cd}{123}${}^m$ $\rightarrow$ \nuc{In}{123}${}^m$ $\rightarrow$ \nuc{Sn}{123}${}^m$. 
The effective BR of $\beta^-$ decay of \nuc{Sn}{123}${}^m$ is estimated to be
99.00\% ($=$57.05\% +26.78\% +0.248\% +14.92\%),
since the $\beta^-$ BR of \nuc{Sn}{123}${}^m$ to \nuc{Sb}{123} is 100\%.

\subsection{$A = 125$ isobars}

There are three decay sequences to \nuc{Sn}{125}${}^m$ from \nuc{Ag}{125}, which has no isomer with a $\beta^-$ branch; 
(a)\nuc{Ag}{125} $\rightarrow$ \nuc{Cd}{125}${}^g$ $\rightarrow$\nuc{In}{125}${}^g$$\rightarrow$\nuc{Sn}{125}${}^m$,
(b)\nuc{Ag}{125}$\rightarrow$\nuc{Cd}{125}${}^g$$\rightarrow$\nuc{In}{125}${}^m$$\rightarrow$\nuc{Sn}{125}${}^m$, and
(c)\nuc{Ag}{125}$\rightarrow$\nuc{Cd}{125}${}^m$$\rightarrow$\nuc{In}{125}${}^g$$\rightarrow$\nuc{Sn}{125}${}^m$.
With the BR of 100\% for the $\beta^-$ decay of \nuc{Sn}{125}${}^m$ to \nuc{Sb}{125},
the effective BR of $\beta^-$ decay of \nuc{Sn}{125}${}^m$ is estimated to be
(89.57 +75*0.1043 $x_{125}$)\%,  which ranges from 89.57\% to 97.39\%,
where $x_{125}$ is the $\beta^-$ BR of \nuc{Ag}{125} to \nuc{Cd}{125}${}^g$ and is not compiled in NUDAT2.

\vspace*{-12pt}

\subsection{$A = 127$ isobars}

There are two decay sequences to \nuc{Sn}{127}${}^m$ from \nuc{Cd}{127}, which has no isomer with a $\beta^-$ branch; 
(a) \nuc{Cd}{127} $\rightarrow$ \nuc{In}{127}${}^g$ $\rightarrow$ \nuc{Sn}{127}${}^m$ and
(b) \nuc{Cd}{127} $\rightarrow$ \nuc{In}{127}${}^m$ $\rightarrow$ \nuc{Sn}{127}${}^m$.
The effective BR of $\beta^-$ decay of \nuc{Sn}{127}${}^m$ is estimated to be (100 -17.17 $x_{127}$)\%, which ranges from 82.83\% to 100\%
with the BR $\beta^-$ decay of 100\% of \nuc{Sn}{127}${}^m$ to \nuc{Sb}{127},
where $x_{127}$ is the $\beta^-$ BR of \nuc{Cd}{127} to \nuc{In}{127}${}^g$ and is not compiled in NUDAT2.
Here we assume that the $\beta^-$ BR of \nuc{Cd}{127}(3/2+) to an another isomer, \nuc{In}{127}${}^{m2}$(21/2-), is negligible 
compared with that of \nuc{Cd}{127}(3/2+) to \nuc{In}{127}${}^{m}$(1/2-) because of the large spin difference between \nuc{Cd}{127}(3/2+) and \nuc{In}{127}${}^{m2}$(21/2-).

\subsection{$A = 128$ isobars}

There is a decay sequence to \nuc{Sb}{128}${}^m$ from \nuc{Cd}{128}, which has no isomer with a $\beta^-$ branch; 
\nuc{Cd}{128} $\rightarrow$ \nuc{In}{128}${}^g$ $\rightarrow$ \nuc{Sn}{128}${}^g$ $\rightarrow$ \nuc{Sb}{128}${}^m$.
The effective BR of $\beta^-$ decay of \nuc{Sb}{128}${}^m$ is estimated to be 96.4\%,
with the $\beta^-$ BR of 96.4\% of \nuc{Sb}{128}${}^m$ to \nuc{Te}{128}.

\subsection{$A = 129$ isobars}

There are two decay sequences to \nuc{Sn}{129}${}^m$ from \nuc{Cd}{129}, which has no isomer with a $\beta^-$ branch;
(a) \nuc{Cd}{129} $\rightarrow$ \nuc{In}{129}${}^g$ $\rightarrow$ \nuc{Sn}{129}${}^m$ $\rightarrow$ \nuc{Sb}{129}${}^m$ and
(b) \nuc{Cd}{129} $\rightarrow$ \nuc{In}{129}${}^m$ $\rightarrow$ \nuc{Sn}{129}${}^m$ $\rightarrow$ \nuc{Sb}{129}${}^m$.
The effective BR of $\beta^-$ decay of \nuc{Sn}{129}${}^m$ is estimated to be
3.74*( 0.002131 + 0.1287 $x_{129}$)*0.85\%, which ranges from 0.068\% to 0.48\%
with the $\beta^-$ BR of 85\% of \nuc{Sn}{129}${}^m$ to \nuc{Sb}{129},
where $x_{129}$ is the $\beta^-$ BR of \nuc{Cd}{129} to \nuc{In}{129}${}^g$ and is not compiled in NUDAT2.
Here the $\beta^-$ BR of \nuc{In}{129} to \nuc{Sn}{129}${}^g$ 
is evaluated to be (10-20)\% in NUDAT2 and is set to be 15\%.

Furthermore, 
we have two branches from \nuc{Sb}{129} to \nuc{Te}{129}${}^m$; 
(a) \nuc{Sb}{129}${}^g$ $\rightarrow$ \nuc{Te}{129}${}^m$, and 
(b) \nuc{Sb}{129}${}^m$ $\rightarrow$ \nuc{Te}{129}${}^m$.
The effective BR of $\beta^-$ decay of \nuc{Te}{129}${}^m$ is estimated to be (6.47-6.6)\%,
since the BR $\beta^-$ decay of \nuc{Te}{129}${}^m$ to \nuc{I}{129} is 37\%.

\subsection{$A = 131$ isobars}

There is a decay sequence to \nuc{Te}{131}${}^m$ from \nuc{Sb}{131}, which has no isomer with a $\beta^-$ branch;
\nuc{Sb}{131} $\rightarrow $ \nuc{Te}{131}${}^m$.
The effective BR of $\beta^-$ decay of \nuc{Te}{131}${}^m$ is estimated to be 4.84\%,
since the BR $\beta^-$ decay of \nuc{Te}{131}${}^m$ to \nuc{I}{131} is 74.5\%.

\subsection{$A = 134$ isobars}

The effective BR of $\beta^-$ decay of \nuc{I}{134}${}^m$ is estimated to be 0\%,
because there is no decay sequences to \nuc{I}{134}${}^m$ from \nuc{Te}{134}, which has no isomer with a $\beta^-$ branch.

\bsp	
\label{lastpage}
\end{document}